\documentclass[journal]{IEEEtran}
\usepackage{stfloats}

\usepackage{enumerate}
\usepackage{algorithm,algpseudocode}
\usepackage{setspace,amsmath,latexsym,cite,amssymb,epsfig,amsfonts}
\usepackage{url,cite}
\usepackage{graphicx}
\usepackage{psfrag}
\usepackage{footmisc}
\usepackage{multirow}
\usepackage{color}
\usepackage{multicol}
\usepackage{mathtools}
\usepackage{subcaption}
\usepackage{graphicx}
\usepackage{amssymb}
\usepackage{amsmath}
\usepackage{epstopdf}
\usepackage{geometry}
\usepackage{float}
\usepackage{balance}

\algnewcommand\algorithmicforeach{\textbf{for}}
\algdef{S}[FOR]{For}[1]{\algorithmicforeach\ #1\ \algorithmicdo}

\allowdisplaybreaks[4]
\twocolumn

\makeatother

        \makeatletter
        \def\fps@eqnfloat{!t}
        \def\ftype@eqnfloat{4}
        
        \newenvironment{eqnfloat*}
               {\@dblfloat{eqnfloat}}
               {\end@dblfloat}
        \makeatother
\title{The Communication and Computation Trade-off in Wireless Semantic Communications}

\author{Xuyang Chen, \IEEEmembership{Student Member, IEEE}, Chong Huang, \IEEEmembership{Member, IEEE}, Gaojie Chen, \IEEEmembership{Senior Member, IEEE}, \\Daquan Feng, \IEEEmembership{Member, IEEE} and Pei Xiao, \IEEEmembership{Senior Member, IEEE}
\thanks{The work presented in this article was supported by the P.R.C. Shenzhen Science and Technology Program under Grant GJHZ20220913143013024, and in part by the Fundamental Research Funds for the Central Universities, Sun Yat-sen University, under Grant No.24hytd010.}
\thanks{X. Chen and D. Feng are with the College of Electronics and Information Engineering, Shenzhen University, Shenzhen 518060, China. Email: chenxuyang2021@email.szu.edu.cn, fdquan@szu.edu.cn.}
\thanks{C. Huang and P. Xiao are with 5GIC \& 6GIC, Institute for Communication Systems, University of Surrey, Guildford, GU2 7XH, United Kingdom, Email: \{chong.huang, p.xiao\}@surrey.ac.uk. (Corresponding author: C. Huang, P. Xiao)}
\thanks{G. Chen is with School of Flexible Electronics (SoFE) \& State Key Laboratory of Optoelectronic Materials and Technologies, Sun Yat-sen University, Guangdong, 510220, China. Email: gaojie.chen@ieee.org.}
}

\begin{document}
\captionsetup[figure]{name={Fig.},labelsep=period}

\begin{singlespace}
\maketitle
\end{singlespace}

\thispagestyle{empty}
\begin{abstract}
Semantic communications have emerged as a crucial research direction for future wireless communication networks. However, as wireless systems become increasingly complex, the demands for computation and communication resources in semantic communications continue to grow rapidly. This paper investigates the trade-off between computation and communication in wireless semantic communications, taking into consideration transmission task delay and performance constraints within the semantic communication framework. We propose a novel trade-off metric to analyze the balance between computation and communication in semantic transmissions and employ the deep reinforcement learning (DRL) algorithm to minimize this metric, thereby reducing the cost associated with balancing computation and communication. Through simulations, we analyze the trade-off between computation and communication and demonstrate the effectiveness of optimizing this trade-off metric.
\end{abstract}

\section{Introduction}
\IEEEPARstart{W}{ith} the rapid development of wireless communications, future sixth-generation (6G) networks need to accommodate the demands of massive user access for the Internet of Things (IoT) and virtual worlds, thereby imposing exceptionally high requirements on wireless transmission rates. To achieve this goal, semantic communications have emerged as one of the key development directions for future wireless networks due to its high spectral efficiency \cite{10183794}. In semantic communications, information is extracted at the transmitter and the extracted semantic information is transmitted to the receiver, then the received semantic information is used to reconstruct the original information at the receiver, significantly reducing the required communication bandwidth. Currently, many research areas have considered the impact of semantic communications in wireless networks, including text transmissions \cite{9763856}, voice transmissions \cite{9450827}, image transmissions \cite{9953076}, non-orthogonal multiple access (NOMA) systems \cite{10158994} and edge computing \cite{10508293}.

Semantic communications can significantly lower the rate requirements for wireless transmission tasks. However, the trade-offs among computation, communication, and reconstruction quality for semantic transmissions are critical in wireless networks \cite{10328186}. A transformer-based reconstruction method for text transmissions was proposed to consider the channel adaptability for semantic communications in \cite{9398576}. To enhance the reconstruction quality in image transmissions, the multi-scale semantic embedding spaces was studied in \cite{10065571} for semantic communications. In \cite{10539255}, the trade-off between communication rate and reconstruction quality was investigated for semantic communications via deep reinforcement learning (DRL) algorithms. The balance between semantic information and transmission security was studied in \cite{10483054}. However, while the balance between compressing the original information and meeting transmission demands in semantic communication has garnered considerable attention, the trade-off between computational demands and communication capabilities in wireless semantic communications remains an unresolved challenge. Moreover, the reconstruction performance at the receiver also influences the trade-off between computation and communication.

In this letter, we investigate the trade-off between communication and computation with reconstruction quality constraint in wireless semantic communications. Our main contributions are summarized as follows:

\begin{itemize}
  \item We define a novel metric, semantic communication cost metric (SCCM), for the trade-off between computation and communication in wireless semantic communications, constrained by the reconstruction quality.

  \item We establish a depth-adaptive semantic communication framework to accommodate the dynamically changing computational and transmission demands of wireless networks. Moreover, to verify the impact of the proposed trade-off metric SCCM in wireless communications, we utilize a DRL algorithm to optimize the resource allocation in the proposed wireless semantic network where channels vary dynamically.

  \item Simulation results analyze the trade-off metric SCCM in wireless semantic communications, and demonstrate the effectiveness of optimizing wireless resource to reduce the cost of semantic communications.
\end{itemize}

The rest of the paper is organized as follows: Section \ref{se:sm} introduces the system model and the problem formulation. The optimization algorithm is presented in Section \ref{se:DRL}. In Section \ref{se:sim}, simulation results are provided to evaluate the trade-off metric. Finally, Section \ref{se:con} concludes the paper.

\section{System Model and Problem Formulation} \label{se:sm}
\subsection{System Model} \label{se:sm1}
In this paper, we consider a wireless network consisting of $M$ base stations (BSs) $S_m$ ($m \in \mathcal M = \{1, 2, ..., M\}$) and $N$ users $U_n$ ($n \in \mathcal N = \{1, 2, ..., N\}$), each BS or user is equipped with a single antenna. We assume $z_{m,n}$ ($z_{m,n} \in \mathbb{Z}$) is a Binary indicator for the user association between $S_m$ and $U_n$, $z_{m,n} = 1$ denotes the BS $S_m$ serves the user $U_n$, otherwise there is no transmission links between $S_m$ and $U_n$. The channel coefficient $h_{m,n}$ between the BS $S_m$ and the user $U_n$ follows Rayleigh fading which remain unchanged during one time slot and vary independently from one time slot to another. Thus, the channel coefficient $h_{m,n}$ can be expressed as $h_{m,n}=g_{m,n} d_{m n}^{\alpha/2}$, where $g_{m,n}$ follows Gaussian distribution with zero-mean and unit-variance, $d_{m,n}$ denotes the distance between the BS $S_m$ and the user $U_n$, $\alpha$ denotes the path loss exponent of Rayleigh fading channels. Therefore, the received signal at $U_n$ is
\begin{equation}\small\label{yUn}
\begin{aligned}
   y_{n} = \sqrt{P} h_{m,n} x_m +  \sum_{v=1,v \neq m}^{M} \sqrt{P} h_{v,n} x_v   + n_{n},
\end{aligned}
\end{equation}
where $P$ denotes the transmit power for BSs, $x_m$ is the signal transmitted from the BS $S_m$, $x_v$ denotes the interference signal from the BS $S_v$, $n_{n}$ is the additive-white-Gaussian-noise (AWGN) with variance ${\sigma}_{n}^2$ at $R_k$. Thus, the transmission rate between $S_m$ and $U_n$ is given by
\begin{equation}\small\label{Capmn}
\begin{aligned}
C_{m,n} = {\rm{log_{2}}} \big(1+ \frac{P |h_{m,n}|^2}{\sum_{v=1,v \neq m}^{M}P |h_{v,n}|^2 + {\sigma}_{n}^2}\big).
\end{aligned}
\end{equation}

In the proposed wireless semantic network, to effectively illustrate the balance between computation and transmission, we simply assume that each base station serves at most one user simultaneously. For the user $U_n$, there are $K_n$ transmission task requests with the arrival modeled based on the FTP model 3, and the transmission delay for each task must not exceed $L_{\rm max}$ time slots. In this study, we assume that all tasks involve the transmission of images. Each BS is equipped with a vision transformer (ViT) \cite{9716741} function to extract semantic information from the original file and subsequently sends the extracted semantic information to the corresponding user. Each user is equipped with a ViT, possessing the necessary knowledge, to reconstruct the corresponding image.

Since semantic information is significantly smaller than the original image, it can greatly reduce the required transmission bandwidth. However, due to the dynamic nature of the communication environment, such as the Rayleigh fading considered in this work, wireless communication links often operate at lower rates than anticipated. This results in increased transmission delays for communication tasks, adversely affecting user experience. Consequently, some existing studies on semantic communication have explored adjusting the semantic information compression ratio by sacrificing a certain level of reconstruction quality to adapt to dynamic channels. However, increasing computational resource consumption can adjust the semantic information compression ratio while ensuring reconstruction quality, thereby reducing the required communication rates and maintaining the transmission quality of semantic communication. Therefore, the trade-off between computation and communication is the primary focus of this work.

\subsection{Semantic Model} \label{se:semantic}
ViT has a powerful capability to capture the complex visual semantics from images and has become a promising tool in the field of image generation. In our semantic communication networks, the ViT-based encoder extracts the semantic information from the original image. The semantic information is then transmitted to the receiver through wireless communication channels. Upon reception, the ViT-based decoder can read the tokens in semantic information to reconstruct the original image.

In this work, we adjust the depth of the ViT-based encoder and decoder in wireless semantic communications to realize the trade-off between communication and computation. The encoder/decoder depth denotes the number of layers in the ViT architecture. By adjusting the depth, we can directly change the computational resources required for processing the semantic information. A deeper encoder can capture more features from the original image, thereby achieving better semantic representations, but it is at the cost of more computational resource. On the other hand, a deeper decoder can reconstruct the original image more accurately based on the semantic information, but this also requires more computational resource. In addition, the number of output symbols generated by the encoder affects the size of the semantic information transmitted to the receiver. Therefore, increasing the depth consumes more computational resource, but this additional computation allows for better extraction of features from the original image, supporting a smaller number of encoder output symbols without degrading the reconstruction quality. In this work, the channel encoder-decoder is implemented using fully-connected neural networks, which facilitate adaptive adjustment of semantic transmission rates through dynamic modulation of output feature dimensions. Thus, we can control computational consumption to adapt to different signal-to-noise ratio (SNR) levels in various communication conditions without compromising the quality of semantic communications.

\begin{figure*}[t!]
  \centering
  \centerline{\includegraphics[width=0.95\textwidth]{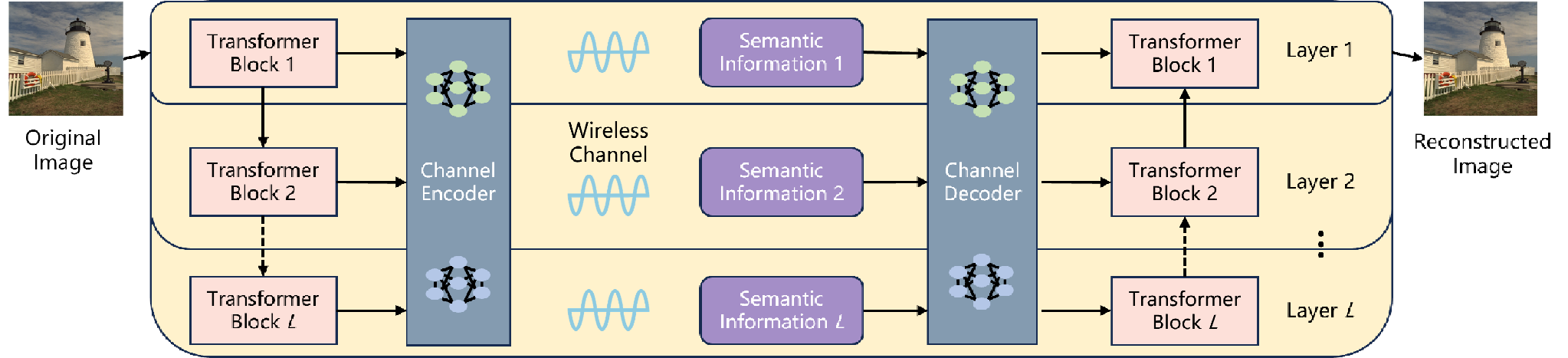}}
 \caption{The structure of the proposed depth-adaptive semantic framework.} \label{fig:semantic}
\end{figure*}

As shown in Fig. \ref{fig:semantic}, the original image is input into a ViT-based encoder in semantic communications. The encoder consists of multiple transformer layers that process the image and output a set of semantic information representing the features of the original image, $I$ denotes the number of total encoder/decoder depth in the semantic framework. This semantic information is then transmitted to the receiver through a wireless communication channel. The received semantic information is input into a ViT-based decoder to reconstruct the image. Both the encoder and decoder are composed of multiple blocks, and the required number of blocks is adjusted based on computational and communication demands. To be specific, more blocks require more computational resource but also enable better extraction of semantic features, thereby reducing the size of the semantic information that needs to be transmitted with minimal loss in performance.

In summary, we train separate semantic communication models for encoder-decoder architectures of different depths, and conduct extensive experiments to identify the most suitable channel codec for each configuration. When selecting an encoder of a particular depth, we switch to the corresponding model rather than extracting intermediate features from a single pre-trained deep model.
\subsection{Problem Formulation} \label{se:problem}
To thoroughly evaluate the performance of semantic communication systems, it is necessary to introduce a new evaluation metric. With the advancement of deep learning technologies, semantic communication systems require substantial computational resources to perform complex semantic analysis and generation. At the same time, to ensure real-time responsiveness and efficiency, the system must efficiently transmit data under limited bandwidth conditions. Traditional performance metrics for semantic communications, such as peak signal-to-noise ratio (PSNR), latency, and bit error rate (BER), are not sufficient to fully describe the trade-off between computation and communication in resource constrained environments. Thus, we design a new metric based on the trade-off between computation and communication to comprehensively assess the performance in semantic information processing and data transmission in wireless networks. The proposed trade-off metric SCCM can be expressed as
\begin{equation}\small\label{metric}
\begin{aligned}
\psi = \varpi_{c} F + \varpi_{t} D,
\end{aligned}
\end{equation}
where $F$ represents the computational consumption measured in giga floating point operations (GFLOPs), $D$ denotes the size of the semantic information transmitted to the receiver, measured in bits. To facilitate the expression of both $F$ and $D$, we normalize them to a range between 0 and 1 according to their respective scales. $\varpi_{c}$ and $\varpi_{t}$ are the weights for computation and communication in the trade-off, respectively, and $\varpi_{c} + \varpi_{t} = 1$. By quantifying the trade-off between computation and communication, SCCM can facilitates the optimization of the resource allocation strategy to enhance user service experience in wireless communications. For example, in scenarios where computational resource are limited but spectrum resource is abundant, the strategy prefers to transmit more semantic information to the receiver. On the other hand, in scenarios where spectrum resource is constrained but computational resource is not, increasing the computational load can reduce the transmitted semantic information size while ensuring the quality of the reconstructed image at the receiver.

To reduce the total resource consumption in wireless semantic communications, we consider optimizing user association and selecting varying depths of encoders/decoders to minimize the sum SCCM $\psi$ in \eqref{metric}. Moreover, we consider the delay and performance constraints in semantic communications. The optimization problem is presented as
\begin{align}
    &\min_{\mathcal{Z}, \mathcal{I}} \sum_{n=1}^{N}\sum_{k=1}^{K_n} \psi_{k} ,\label{SecrecyFunc}\\
    {\rm s.t.}& \sum_{m=1}^{M} z_{m,n} \leq 1,~\forall n \in {\mathcal N} \tag{\ref{SecrecyFunc}{a}}, \label{SecrecyFuncSuba}\\
    &\sum_{n=1}^{N} z_{m,n} \leq 1,~\forall m \in {\mathcal M} \tag{\ref{SecrecyFunc}{b}}, \label{SecrecyFuncSubb}\\
    &L_k \leq L_{\rm max}  \tag{\ref{SecrecyFunc}{c}}, \label{SecrecyFuncSubc}\\
    &PSNR_k \geq PSNR_{\rm min} \tag{\ref{SecrecyFunc}{d}}, \label{SecrecyFuncSubd}\\
    &\mathcal{Z} = \{\mathbb{Z}(1), \mathbb{Z}(2), ..., \mathbb{Z}(T)\} \tag{\ref{SecrecyFunc}{e}}, \label{SecrecyFuncSube}\\
    &\mathcal{I} = \{\mathbb{I}(1), \mathbb{I}(2), ..., \mathbb{I}(T)\} \tag{\ref{SecrecyFunc}{f}}, \label{SecrecyFuncSubf}
\end{align}
where $\mathbb{I}$ denote that at time slot $t$ the depths of encoders/decoders in the semantic communications framework as in Fig. \ref{fig:semantic}, $\mathbb{Z}(t)$ denotes the Binary indicator for the user association at time slot $t$, $t \in \{1, 2, ..., T\}$, $T$ denotes the total time slot for completing all tasks. $L_k = t_{k,f} - t_{k,s}$ denotes the delay of the $k$-th task, where $t_{k,f}$ denotes the time slot that $k$-th transmission task is finished, $t_{k,s}$ denotes the time slot that $k$-th task arrives at BSs. \eqref{SecrecyFuncSuba} and \eqref{SecrecyFuncSubb} shows that one BS can only serves one user, and one user can only access to one BS. \eqref{SecrecyFuncSubc} and \eqref{SecrecyFuncSubd} presents the delay and reconstruction quality constraints for all tasks, respectively, where $PSNR_{\rm min}$ is the reconstruction quality threshold. \eqref{SecrecyFuncSube} and \eqref{SecrecyFuncSubf} indicates the user association and depths of encoders/decoders, respectively. We aim to minimize the sum cost of semantic communication tasks, where each task has delay and reconstruction quality constraints as in \eqref{SecrecyFuncSubc} and \eqref{SecrecyFuncSubd}. Considering the time-varying Rayleigh fading wireless environment, task delay constraints, and the arrival of tasks at non-fixed time slots, this is a long-term optimization issue. To address this problem, we employ a DRL algorithm to adaptively learn from the dynamic wireless environment and accommodate different trade-off weights. This approach provides an adaptable trade-off analysis scheme for semantic communications across various wireless conditions.

\section{DRL-Based Optimization Algorithm}\label{se:DRL}
DRL is well-suited for solving long-term optimization problems and can adapt to dynamically changing wireless communication environments after training in various settings. To utilize DRL, we first formulate the problem as a Markov Decision Process (MDP). In the MDP, the state is used to describe the variables of the wireless communication environment as $s(t) = \{\mathcal{H}(t), \mathcal{K}(t), \mathcal{L}(t), \mathcal{F}(t), \mathcal{D}(t)\}$, where $\mathcal{H}(t)$ is the set of all channel coefficients between BSs and users at time slot $t$, follows block fading based on \eqref{Capmn}. $\mathcal{K}(t)$ denotes the set of the transmission status of arrived tasks at time slot $t$, it follows FTP model 3 \cite{3GPPFTPmodel3} to generate new arrivals, $\mathcal{L}(t)$ denotes the set of delay of arrived tasks, $\mathcal{F}(t)$ and $\mathcal{D}(t)$ denotes the computational cost and communication cost at time slot $t$, respectively. Then, we define the action at time slot $t$ as $a(t) = \{\mathbb{Z}(t), \mathbb{I}(t)\}$. Action is used to control the optimization variables in the problem formulation \eqref{SecrecyFunc}. The reward function is defined as
\begin{equation}\small\label{eq:reward}
\begin{aligned}
    r(t) = - \sum_{n=1}^{N}\sum_{k=1}^{K_n} (\psi_{k} - [L_k - L_{\rm max}]^+ - [PSNR_{\rm min} - PSNR]^+),
\end{aligned}
\end{equation}
where $[x]^+$ = max$(0, x)$. The negative cost is utilized to guide the DRL agent in minimizing the cost and guarantee the constraints in \eqref{SecrecyFunc}, thereby achieving the trade-off between computation and communications. In this work, we introduce soft actor-critic (SAC) algorithm to solve the proposed issue. Since DRL algorithms such as SAC have been extensively applied to wireless communications, we will provide a brief overview of the principle of SAC \cite{haarnoja2018soft}. We can minimize the soft Bellman residual to update the soft-Q value in SAC
\begin{equation}\small\label{eq:SoftQFunction}
\begin{aligned}
W_{Q_{\omega}} = {\mathbb E}_{\big(s(t), a(t)\big) \sim R} \bigg[ \frac{1}{2} \Big(Q_{\omega}\big(s(t), a(t)\big) - \bar{Q}\big(s(t), a(t)\big) \Big)^2  \bigg],
\end{aligned}
\end{equation}
where $Q_{\cdot}$ denotes the value in SAC, ${\mathbb E}[\cdot]$ is the expectation, $R$ denotes the training experience distribution, $\bar{Q}(s(t), a(t)) = r(s(t), a(t)) + \gamma{\mathbb E}_{s(t+1) \sim \rho} [Q_{\bar{\kappa}}(s(t+1))]$, $\gamma$ presents the discount factor, $\rho$ is the transition probability. Thus, we could obtain the gradients as
\begin{equation}\small\label{eq:SoftQGradient}
\begin{aligned}
\bar{\tau}_{\kappa} W_{Q_{\omega}} = &~{\tau}_{\kappa}Q_{\omega}\big(s(t), a(t)\big) \Big(Q_{\omega}\big(s(t), a(t)\big) - r\big(s(t), a(t)\big)\\
                                        &- \gamma Q_{\bar{\kappa}}\big(s(t+1)\big)\Big).
\end{aligned}
\end{equation}
On the other hand, we rebuild the policy network with Gaussian noise and update the policy network $\upsilon$ as
\begin{equation}\small\label{eq:PolicyFunctionNoise}
\begin{aligned}
W_{Q_{\upsilon}} = &~{\mathbb E}_{s(t) \sim D, \varsigma(t) \sim \phi}\bigg[{\rm log}Q_{\upsilon}(f_{\upsilon}\big(\varsigma(t); s(t))|s(t)\big)\\
                    &- Q_{\omega}\big(s(t), f_{\upsilon}(\varsigma(t); s(t))\big) \bigg].
\end{aligned}
\end{equation}
Therefore, we can obtain the gradient of the policy as
\begin{equation}\small\label{eq:PolicyFunctionGradient}
\begin{aligned}
\bar{\tau}_{\kappa} W_{Q_{\upsilon}} = &~{\tau}_{\upsilon}{\rm log}Q_{\upsilon}\big(a(t)|s(t)\big) + \bigg({\tau}_{a(t)}{\rm log}Q_{\upsilon}\big(a(t)|s(t)\big)\\
                                           &- {\tau}_{a(t)}Q\big(s(t), a(t)\big) \bigg) {\tau}_{\upsilon}f_{\upsilon}(\varsigma(t); s(t)).
\end{aligned}
\end{equation}

One of the advantages of the proposed method is its adaptability to dynamic environments, which is critical for real-time wireless communication systems, this method allows for better utilization of resources, especially in computing and bandwidth resource constrained scenarios. Another important advantage is the flexibility of the reward structure in DRL, which enables the system to incorporate diverse objectives, such as minimizing energy consumption, maximizing task throughput, or maintaining a balance between communication and computation costs.

\section{Numerical Results}\label{se:sim}
In the simulation, we utilized Nvidia A100 and PyTorch to train the proposed semantic framework. Unless otherwise stated, the simulation parameters are set as follows: the transmit power gain $P/{\sigma}_{n}^2 = 5$ dB, the path loss exponents $\alpha = 3$. The number of BSs $M = 3$, the number of users $N = 3$, the locations of BSs and users are randomly distributed in a square of 50 m $\times$ 50 m, the bandwidth $B = 10$ MHz, the size of original image for all tasks is 50 KB, the depth of ViT are set as $\{1, 2, 3, 4, 5, 6\}$, the delay and performance thresholds $L_{\rm max} = 3$, $PSNR_{\rm min} = 30$, the number of semantic transmission tasks at each user is 100.

\begin{figure}[t!]
  \centering
  \centerline{\includegraphics[scale=0.5]{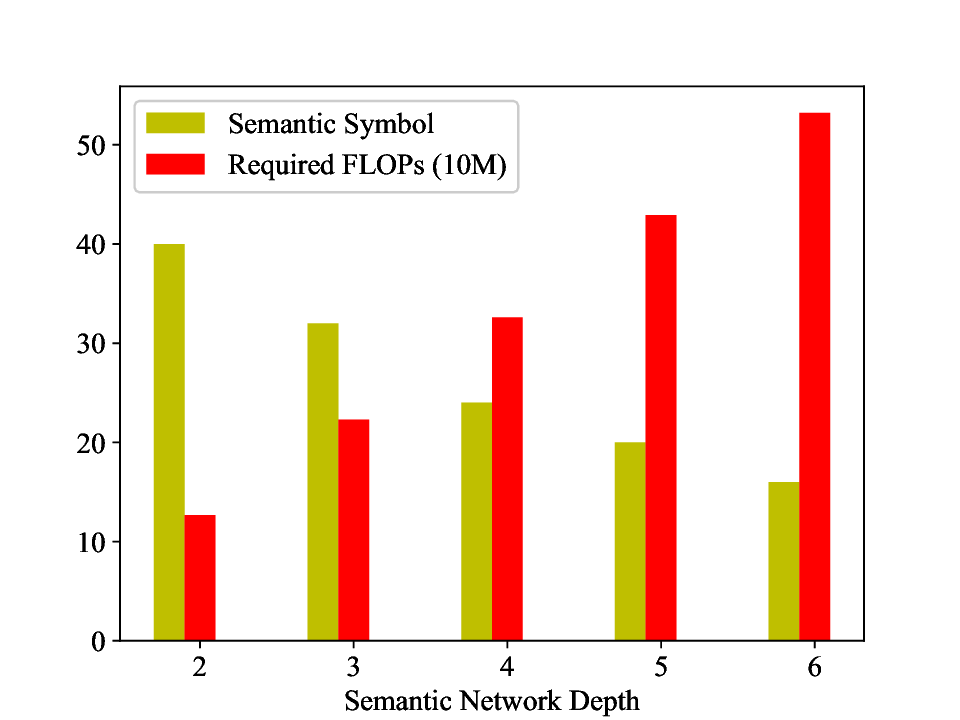}}
 \caption{The comparison between computation and communication in the proposed semantic network.} \label{fig:R1}
\end{figure}
Fig. \ref{fig:R1} presents a comparison between computation and communication in the proposed network. The semantic symbols denote the output vectors from the semantic encoder, FLOP is used to measure the required computational resource. As shown in this figure, the required computation resource of a task rises with an increase in the depth of the semantic network. This is due to the use of deeper networks for compressing and parsing semantic information. Although deeper networks require more computational resource to operate, they are able to capture and form high-quality semantic representations, allowing for a reduction in the size of semantic information while maintaining performance in the dynamic wireless communication networks.

\begin{figure}[t!]
  \centering
  \centerline{\includegraphics[scale=0.5]{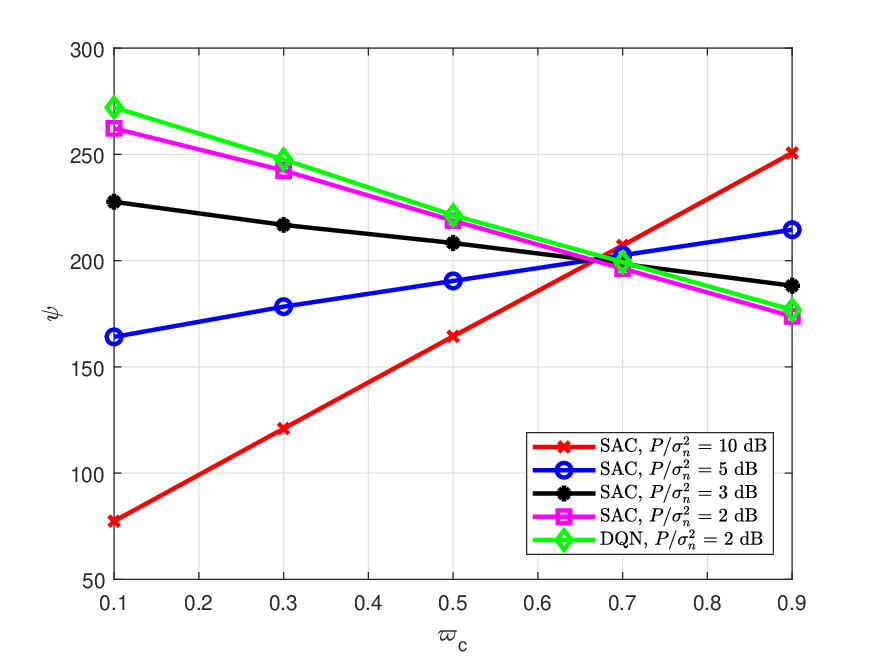}}
 \caption{Sum SCCM versus the weight of computation $\varpi_c$.} \label{fig:R2}
\end{figure}
As illustrated in Fig. \ref{fig:R2}, we compare the sum SCCM to show the impact and trade-off between computation and communication in wireless semantic networks. This figure illustrates that when wireless transmission resources are abundant, it is possible to maximize the utilization of transmission resources to reduce reliance on computation with a power gain of 10 dB. Therefore, a smaller computation weight leads to a lower sum SCCM. Conversely, in scenarios where spectrum resources are scarce, such as a power gain of 2 dB, computational resources are utilized to compress the semantic information to conserve spectrum resource. In such cases, a larger weight should be assigned to computation. The reason is, when the optimization selecting deeper encoder/decoder to reduce the size of semantic information transmitted, the computational cost increases and need to minimizing its costs. Conversely, when a shallower encoder/decoder configuration and larger transmitted semantic information is selected, the algorithm focuses on reducing communication costs. This trade-off is particularly clear in scenarios with constrained resources, such as limited computational power or spectrum availability. Moreover, the figure presents that the proposed SAC-based optimization achieves better result than that from deep Q-Network (DQN), this verifies the advantage of using SAC in such optimization problems.

\section{Conclusion}\label{se:con}
This paper analyzes the trade-off between computation and communication in wireless semantic communications. By adjusting the depth of the encoder and decoder, we can reduce the size of semantic information required for transmission by increasing computational cost, while ensuring the performance of the reconstructed data. Conversely, we can increase the size of the semantic information to lower the computational requirements, while still maintaining the performance of the reconstructed data. Moreover, by joint optimizing user association in wireless communications and the depth of the encoder and decoder in semantic communications, we conduct a comprehensive analysis of the trade-off between computation and communication. Simulation results demonstrate that in dynamically changing wireless communication environments, the trade-off metric SCCM can play a key role in reflecting the resource cost of semantic communications. Future research can further explore trade-offs between computation and communication in multi-user and multi-task scenarios for advancing wireless semantic communication networks.


\bibliographystyle{ieeetr}
\bibliography{ref}


\end{document}